\magnification=\magstep1
\hfuzz=6pt
\baselineskip=18pt

$ $

\vskip 1in

\centerline{\bf Quantum enigma machines}

\bigskip

\centerline{Seth Lloyd}

\centerline{Department of Mechanical Engineering}

\centerline{Massachusetts Institute of Technology}

\centerline{MIT 3-160, Cambridge MA, 02139, USA}

\bigskip\noindent{\it Abstract:}  
Enigma machines are devices that perform cryptography using 
pseudo-random numbers.   The original enigma
machine code was broken by detecting hidden patterns in these 
pseudo-random numbers.   
This paper proposes a model for a quantum optical enigma machine and 
shows that the phenomenon
of quantum data locking makes such quantum enigma machines provably 
secure even in the presence of noise and loss.

\vskip 1cm

The enigma machine used for cryptography during the second world war
was a device which, given a short
keyword, produced a pseudorandom output [1-3] which could be decoded
by a second machine using the same keyword.
The original enigma machine consisted of a series of rotors through
which electrical current could pass in a way that depended on the
relative orientation of the rotors.   The path taken by the current
connected an input symbol to an output system.  After each key press
the rotors went through a stepping motion that changed the
functional relationship betweein input and output for the next
key press.  A sender and receiver who prepared their machines
using the same initial setting, determined by the keyword,
could then exchange encrypted messages.  While the enigma
machine did a pretty good job of scrambling the input, the
outputs deviated sufficiently from pseudorandom sequences
that the enigma code could be broken.  In general, 
classical codes based on pseudo-random  
numbers are secure only if $P \neq NP$ [1-3]: proving the security
of such codes is accordingly difficult.  
This paper proposes a quantum optical version of 
the enigma machine and shows that it is secure in principle,
in the sense that amount of information that Eve can access
about the message can be made arbitrarily small,
even in the presence of arbitrary amounts of loss.

The security of quantum enigma machines relies on
the phenomenon of quantum data locking [4-10].
Suppose that Alice
possesses an $n$-bit message $j$ that she wishes to send to
Bob.   Alice and Bob initially
possess a secret, fully random $m$-bit string $k$ (the `seed'),
where $m<<n$.  They publicly agree upon a set of $2^m$
unitary operations $U_k$, randomly selected according to
the Haar measure.  
Alice first maps the message $j$ to a quantum state $|j\rangle$.
She then applies the transformation $U_k$ corresponding
to the shared seed $k$ and sends the resulting state $|j\rangle_k = 
U_k|j\rangle$ to Bob.  Bob decodes the message by applying the
inverse transformation $U^\dagger_k |j\rangle_k = |j\rangle$.
The devices that perform Alice's and Bob's encoding and decoding
operations can be termed quantum enigma machines, in analogue to 
classical enigma machines that encode and decode via classical 
invertible transformations.  

The method of quantum data locking [4-10] can now be
used to show that 
quantum enigma machines are secure against
eavesdropping.  Suppose that Eve intercepts the full state 
sent by Alice.
Assume that all $2^n$ messages that Alice could send are equally
likely (as is the case, for example, if Alice has performed
good classical data compression on her original message).
Since Eve does not know the seed $k$, the state that she
receives is 
$$2^{-m-n} \sum_{jk} U_k |j\rangle\langle j| U^\dagger_k
= 2^{-m-n} \sum_{jk} |j\rangle_k\langle j|.\eqno(1)$$
The maximum amount of information that Eve can obtain about
Alice's message is then limited by the accessible information $I_c$, 
equal to the maximum mutual information between the inputs
$j$ and the outcome of a measurement made by Eve on the encoded
state.  In [4], it is shown by a simple convexity argument
that the accessible information in turn is limited by
$$I_c \leq n + 2^{-m} \max_{|\phi\rangle}
\sum_{jk} |\langle \phi | j\rangle_k|^2 
\log |\langle \phi | j\rangle_k|^2.\eqno(2)$$
Hayden {\it et al.} [5] have shown that if the $U_k$ are fully
random, i.e., selected from the set of all $n$-qubit unitaries
according to the Haar measure, then Eve's accessible
information can be made arbitrarily small using a key
of size $ m = O(\log n)$.  A number of papers
have extended this result [6-10].  In particular, as
shown in [10], if $m= O(4\log(1/\epsilon))$, then for any $\epsilon >0$
there exists an $n$ sufficiently large that Eve's accessible information
is lower than $\epsilon n$.  
Bob can decode the full $n$-bit message, while in the absence
of those $m$ bits, Eve can obtain only an arbitrarily small amount
of information.  For example, if $\epsilon = n/2^{10}$,
then a key of length $m = O(4\log n + 40)$ allows
Eve access to less than a thousandth of a bit. 
Quantum enigma machines are 
secure in principle by the accessible information criterion.  

There are several hurdles to overcome in order to make quantum
enigma machines secure in practice.
Quantum data locking hides large amounts of data with a small key. 
Consequently, Alice and Bob must be very careful to ensure the security
of that key.  Data locking is susceptible to plain-text attack: if Eve
knows part of the message, she can use that to determine the key and
unlock the rest [7,9].  Alice and Bob can evade the plain-text attack
by using data locking to distribute a random secret key [11]:
Alice simply sends Bob a random number, known only to her.    
Performing a fully random unitary is computationally
inefficient: in [10], however, it is shown how locking can
be performed using non-random unitaries that can be constructed
in time almost linear in $n$.  That is, unlike the classical pseudo-random
transformations of the original enigma machine, quantum pseudo-random
transformations perform a sufficiently good job of scrambling the message
that Eve cannot decipher it.  Perhaps the most pressing issue, however, is the
ability of quantum enigma machines to function effectively in
the presence of noise and loss.

\bigskip\noindent{\it Noise and Loss}

Bob's ability to decode the message is sensitive
to noise and loss on the channel.  As stated, the protocol
requires a communications channel capable of sending quantum
information, which suggests that more than $50\%$ loss would
render the protocol ineffective.  I now show that, contrary
to that intuition, quantum enigma machines can function with
arbitrarily high levels of loss.   

First consider the depolarizing channel.  Alice sends qudits
$\in C^d$ to Bob.  With probability $\eta$ the qudit is transmitted
faithfully, and with probability $1-\eta$ it is replaced with
a fully mixed state.  For $\eta \leq 1/2 $, this channel is
anti-degradable and has neither
quantum capacity nor private capacity [12-13].  The enigmatic version
of the depolarizing channel is straightforward: Alice applies
a random unitary to the qudit before sending, and Bob applies
the inverse transformation on the other side.  To Alice and Bob,
then, the channel behaves like the ordinary depolarizing channel,
which has classical capacity for all $\eta > 0$.
By contrast, even if Eve intercepts the locked qudit in a noiseless
state, she can extract only a vanishingly small amount of accessible
information. 

To send information securely by the accessible information criterion,
Alice and Bob agree on a particular error correcting
code for sending classical information down the depolarizing
channel.  The code has block length $b$ appropriate for
the depolarizing rate $\eta$ and the degree of accuracy
that they wish to attain.  As shown in [10], if a single
use of the locked channel bounds Eve's accessible information
by $\epsilon$, then the composition
of $b$ uses of a locked channel bounds her accessible information
by $b\epsilon$.  For sufficiently large $d$, then, Alice and
Bob can pick a key of length $O( b \log(b/\epsilon))$ to lock
the $b$ uses of the channel, allowing Eve access to at
most a fraction $\epsilon$ of the transmitted information.
 
The example of the qudit depolarizing channel shows that quantum
data locking combined with suitable error correction can
still send information that is secure by the accessible
information criterion down a quantum channel whose
private capacity is zero.  Now turn to an example
of how quantum data-locking can be used in an experimentally
achievable setting, even in the presence of large amounts of loss.
Consider the lossy bosonic channel [14-16].
Alice's and Bob's quantum enigma machines can consist
of passive linear elements such as beam splitters and phase
shifters.  Alice's quantum enigma machine maps the annihilation
operators for $N$ input modes $\vec \alpha \rightarrow 
U_k  \vec \alpha$, where $U_k$ is an $N\times N$  random unitary
transformation acting on the mode labels, selected from a 
set of $M$ such transformations.  Bob's enigma
machine performs the inverse
transformation, $\vec \alpha \rightarrow 
U^\dagger_k  \vec \alpha$.  Assume for the moment that there
is no loss in the encoding and decoding procedures -- all
the loss is in the communication channel.  

To see that Alice's and Bob's enigma machines can still perform
effectively in the presence of high levels of loss, consider
a `unary' encoding, in which $ n = \log_2 N$ message bits are
encoded on a single photon spread amongst $N$ modes, so
that the $j$'th message is encoded in the state $|j\rangle$
with the single photon in mode $j$.  The security analysis
for this unary encoding is mathematically the same as the
qubit case above.  Alice's quantum enigma machine uses linear
optics to apply a random unitary transformation $U_k$ to
the mode labels, yielding the state $|j\rangle_k = U_k|j\rangle$.
In the absence
of knowledge of the secret key $k$, Eve's accessible
information about the message can be made less than
$\epsilon$ bits using a key of size $O(\log n \log(n/\epsilon))$.

A simple strategy now allows Alice and Bob to communicate
even in the presence of high amounts of loss. 
Since Bob's quantum enigma machine undoes
Alice's linear transformation, if the photon that Alice sent
does shows up, Bob's enigma machine maps it into the $j$'th mode.    
If no photon shows up, Bob simply asks Alice to resend with
new key.
If $\epsilon$ is the fraction of the information that Eve has
access to for one transmission, then $T\epsilon$ is
the fraction that she can obtain after $T$ transmissions.
For a fixed level of loss per mode, 
the probability of losing the photon is independent
of the number of modes $N$.  Consequently,  
by making $\epsilon$ small and $N$ large,
Alice and Bob can cope with arbitrily high levels of loss while guaranteeing
that Eve gains an arbitrarily small fraction of the transmitted
information.

The ability of Bob to unlock information even in the presence of
large amounts of loss comes about because the lossy bosonic channel
retains quantum capacity with feedback 
up to arbitrarily high loss levels.  The non-trivial
feature of this quantum enigma machine is that the locking/unlocking protocol 
survives the transmission/feedback process.  
Since unary coding over blocks of $N$ modes attains the
channel capacity for a lossy channel in the limit of low photon
number per mode [14-16], the capacity of a secure quantum enigma machine
channel is essentially the same as that of an insecure bosonic
channel in this regime.  

In the presence of thermal noise and loss, the capacity of the
bosonic quantum enigma machine is bounded below by the forward and 
reverse coherent information [17-18].    
At optical frequencies, the background thermal
noise is neglible.  Moreover, the conventional way for Alice to
generate her single photons is by parametric downconversion of a
pair of photons: detection of one photon of the pair then heralds
the arrival of the other.   To jam the channel, Eve must introduce
enough photons to insure that the exact time bin of the heralded
photon contains excess photons.  As long as the average noise
photon number within the time bin is 
below the threshold for non-zero forward or reverse coherent information,
Alice and Bob retain secret capacity via quantum locking, both for
direct transmission (with error-correcting codes) and for key
generation (with privacy amplification).  

\bigskip\noindent{\it Coherent states}

An important open question is whether it is possible to construct
a provably secure quantum enigma machine using linear optics and 
coherent states.
An example of a high-power coherent-state quantum enigma machine 
is Yuen's $\alpha$-$\eta$ model [19-23].  
In the quantum enigma model discussed here, the 
transformations $U_k$ map the message to sets of overlapping
coherent states is selected at random, whereas in the  $\alpha$-$\eta$ model
these transformations are selected in a systematic fashion using
binary phase shift keying [19].  The $\alpha$-$\eta$
model has been shown to yield error probabilities approaching
1 for common types of eavesdropper [19-23], although its 
security based on accessible information or on
Holevo information [24-25] has not been shown.

\bigskip\noindent{\it Conclusions:}

This paper showed that quantum enigma machines, unlike their
classical counterparts, are provably secure against eavesdropping.
The security of the quantum enigma machine is guaranteed by
its ability to spread encoded states over Hilbert space via
quantum data locking,
thereby limiting the ability of an eavesdropper to obtain
information about the encoded message.  A quantum enigma machine
based on single photons and unary encoding was exhibited and shown to
retain its security in the presence of noise and high amounts of loss in
the communication channel.  The general question of how to define
the capacity of quantum enigma machines that render a specified
channel secure  by the accessible information
remains open [26].  
Here we exhibited enigmatic quantum coding schemes for the depolarizing
and lossy channels: it would be useful to have codes for quantum data 
locking on general quantum channels. 
As with any cryptographic system,
the security of practically realizable quantum
enigma machines must be investigated on a case-by-case basis to
probe for possible attacks.    

\vfill \eject
\noindent{\it Acknowlegments:} This work was supported by DARPA.
The author would like to thank C. Lupo, V. Giovannetti, S. Guha, P. Kumar,
J.H. Shapiro,  M. Takeoka, M. Wilde, and H. Yuen
for helpful discussions.

\vskip 1in

\noindent{\it References:} 

\bigskip\noindent [1] W. Diffie, M.E. Hellman, {\it IEEE Trans. Inf. Th.} {\bf IT-22}
644-655 (1976). 

\bigskip\noindent [2] D.R. Stinson, {\it Cryptography: theory and practice},
Chapman and Hall, Boca Raton, 2006.

\bigskip\noindent [3] A.A. Bruen, M.A. Forcinto, {\it Cryptography, Information Theory, and Error-Correction: A Handbook for the 21st Century},  J. Wiley, Hoboken (2006).

\bigskip\noindent [4] D.P. DiVincenzo, M. Horodecki, D.W. Leung, J.A. Smolin,
B.A. Terhal, {\it Phys. Rev. Lett.} {\bf 92}, 067902 (2004);
arXiv: quant-ph/0303088.

\bigskip\noindent [5] P. Hayden, D. Leung, P.W. Shor, A. Winter, 
{\it Com. Math. Phys.} {\bf 250}, 371-391 (2004); arXiv: quant-ph/0307104.

\bigskip\noindent [6] H. Buhrman, M. Christandl, P. Hayden, H-K Lo,
S. Wehner, {\it Phys. Rev. A} {\bf 78}, 022316 (2008); arXiv: quant-ph/0504078.

\bigskip\noindent [7] R. K\"onig, R. Renner, A. Bariska, U. Maurer,
{\it Phys. Rev. Lett.} {\bf 98}, 140502 (2007).
arXiv: quant-ph/0512021.

\bigskip\noindent [8] D. Leung, {\it International Workshop on Statistical-Mechanical Informatics 2008 (IW-SMI 2008)}, {\it J. Phys.: Conference Series} {\bf 143},
012008 (2009).

\bigskip\noindent [9] F. Dupuis, J. Florjanczyk, P. Hayden, D. Leung, 
`Locking classical information,'
arXiv: 1011.1612.

\bigskip\noindent [10] O. Fawzi, P. Hayden, P. Sen, 
`From low-distortion norm embeddings to explicit uncertainty relations 
and efficient information locking,' arXiv: 1010.3007.

\bigskip\noindent [11] U.M. Maurer, {\it IEEE Trans. Inf. Th.} {\bf 39},
733 (1993).

\bigskip\noindent [12] G. Smith, J.A. Smolin,
Proceedings of the IEEE Information Theory Workshop 2008. pp 368-372:
arXiv: 0712.2471. 

%\bigskip\noindent [13] C. Morgan, A. Winter,
%` ``Pretty strong" converse for the quantum capacity of degradable channels,'
%arXiv: 1301.4927.

\bigskip\noindent [13] F.G.S.L. Brandao, J. Oppenheim, S. Strelchuk,
{\it Phys. Rev. Lett.} {\bf 108}, 040501 (2012); arXiv: 1107.4385. 

\bigskip\noindent [14] Caves, P.D. Drummond, {\it Rev. Mod. Phys.} {\bf 66}, 481-537
(1994).

\bigskip\noindent [15] V. Giovannetti, S. Guha, S. Lloyd, L. Maccone, J.H. Shapiro, H.P.
Yuen,  {\it Phys. Rev. Lett.} {\bf 92}, 027902 (2004).

\bigskip\noindent [16] V. Giovannetti, S. Guha, S. Lloyd, L. Maccone, J.H. Shapiro,
{\it Phys. Rev. A} {\bf 70}, 032315 (2004); arXiv:quant-ph/0404005.

\bigskip\noindent [17] R. Garcia-Patron, S. Pirandola, S. Lloyd, J.H.
Shapiro, {\it Phys. Rev. Lett.} {\bf 102}, 210501 (2009);  arXiv:0808.0210.

\bigskip\noindent [18] S. Pirandola, R. Garcia-Patron, S.L. Braunstein,
S. Lloyd, {\it Phys. Rev. Lett.} {\bf 102}, 050503 (2009); arXiv: 0809.3273.

\bigskip\noindent [19] H.P.Yuen, {\it Proc. QCMC’00, Capri,} P.Tombesi and O.Hirota eds., 
Plenum Press, 2001.

\bigskip\noindent [20] G.A. Barbosa, E. Corndorf, P. Kumar, H.P. Yuen,
{\it Phys. Rev. Lett.} {\bf 90}  227901 (2003);  arXiv: quant-ph/0212018

\bigskip\noindent [21] E. Corndorf, G. Barbosa, C. Liang, H.P. Yuen, P. Kumar,
{\it Optics Letters} {\bf 28}, 2040-2042 (2003).

\bigskip\noindent [22] O.Hirota, K.Kato, M.Sohma, T.Usuda, and K.Harasawa,
{\it Proc. Q. Commun. Q. Imaging, SPIE} {\bf 5551}, (2004);
arXiv: quant-ph/0407062.

\bigskip\noindent [23] O. Hirota, M. Sohma, M. Fuse, K. Kato, {\it Phys. Rev. A}
{\bf 72}, 022335 (2005);
arxiv: quant-ph/0507043.

\bigskip\noindent [24] I. Devetak, A. Winter, 
{\it Proc. R. Soc. Lond. A} {\bf 461}, 207-235, (2005);  
arXiv: quant-ph/0306078.

\bigskip\noindent [25] I. Devetak, {\it IEEE Trans. Inf. Theory} {\bf 51},
44 (2005).

\bigskip\noindent [26] C. Lupo, S. Guha, P. Hayden, H. Krovi, 
S. Lloyd, J. H. Shapiro, M. Takeoka, and M. M. Wilde, `Quantum 
enigma machines and the locking capacity of a quantum channel.' In preparation.

\vfill\eject\end

\bigskip\noindent [9] O. Hirota, {\it Phys. Rev. A} {\bf 76}, 032307 (2007).

\bigskip\noindent [17] J. Emerson, Y.S. Weinstein, M. Saraceno,
S. Lloyd, D.G. Cory, {\it Science} {\bf 302}, 2098 (2003).

\bigskip\noindent [18]  J. Emerson, E. Livine, S. Lloyd,  
{\it Phys. Rev. A} {\bf 72}, 060302(R) (2005).

\bigskip\noindent [20] A. Tomita, O. Hirota, {\it J. Opt. B} {\bf 2},
705 (2000).

\bigskip\noindent [21] A.D. Wyner, {\it Bell Sys. Tech. J.} {\bf 54},
1355 (1975).

\bigskip\noindent [22] A. Khisti, G. Wornell, A. Wiesel, Y. Eldar,
{\it IEEE International Symposium on Information Theory}, 2471-2475
(2007).

\bigskip\noindent [23] C.H. Bennett, G. Brassard, {\it Proceedings
of IEEE International Conference on Computers, Systems and Signal 
Processing, Bangalore, India, 1984}, (IEEE Press, 1984), pp. 175–179.

\bigskip\noindent [24] A. Ekert, {\it Phys. Rev. Lett.} {\bf 67},
661-663 (1991).

\bigskip\noindent [25] P.W. Shor, J. Preskill, {\it Phys. Rev. Lett.}
{\bf 85}, 441-444 (2000).

\vfill\eject\end